\documentclass{revtex4}
\usepackage{epsf,graphicx}
\usepackage{amsmath,amssymb,amsfonts}
\newcommand{\nvec}[1]{\stackrel{\rightarrow}{#1}}
\begin{document}

\author{D.H.E.~Gross}

\affiliation{{$^1$}Hahn-Meitner-Institut, Bereich Theoretische
Physik, Glienickerstr. 100, 14109 Berlin (Germany)}

\title{The microcanonical entropy is multiply differentiable\\No dinosaurs in microcanonical gravitation:\\
No special "microcanonical phase transitions"}

\begin{abstract} The microcanonical entropy $S=\ln[W(E)]$ is the {\em geometrical measure}
of the microscopic redundancy or ignorance about the N-body system. {\em
Even for astronomical large systems} is the microcanonical entropy
everywhere {\em single valued and multiply differentiable}. Also the
microcanonical temperature  is at all energies single valued and
differentiable. It is further shown that the recently introduced
singularities of the {\em microcanonical} entropy like "microcanonical
phase transitions", and exotic patterns of the {\em microcanonical} caloric
curve $T(E)$ like multi-valuednes or the appearance of "dinosaur's necks"
are inconsistent with Boltzmann's fundamental definition of entropy.

\end{abstract}
\maketitle
\section{Introduction}
In a recent paper Barr\'{e}\cite{bouchet03} introduced a special class of
"microcanonical" phase transitions. These are defined by singularities of
the microcanonical entropy or the microcanonical temperature.  However,
such singularities do not exist in the microcanonical thermodynamics based
on Boltzmann's principle \cite{gross174} neither for small systems nor for
the really large self-gravitating ones.

We will show in the next section, the entropy as defined by Boltzmann's
principle
\begin{equation}S(E)=\ln[W(E)]\end{equation}
as also the inverse temperature
\begin{equation}\beta(E)=T^{-1}(E)=\frac{d S}{dE}>0\end{equation}
are single valued, smooth, and multiple differentiable ($\frac{3N-7}{2}$
times) for any finite,  especially so for astronomically large $N$. This,
obviously, means that $S(E)$ as well $\beta(E)$, and consequently
$T(E)=1/\beta$ are everywhere {\em single} valued and multiply
differentiable.

Here $W(E)$ is the ($6N-1$) folded integral i.e. the {\em
geometric area} of the microcanonical many-fold of points with the
energy $E$ in the $6N$-dim. phase space. It measures the number
(in units of $2\pi\hbar$) of points in the $6N$-dim phase space
which are consistent with all known and dynamically conserved
constraints of the system. If no other conservation law, i.e. time
independent constrain, exists, the usual assumption is, the
phase-space point of the system moves as time passes over the
whole remaining sub-manifold (here the microcanonical energy
shell) of the $6N$-dim. phase space. The dynamics is "mixing". In
some cases like nuclear collisions, mixing may even not be needed.
As we don't have any further control every point of the manifold
can be the initial point of the system. Here we assume with
Boltzmann, there is no preference of any initial region of the
allowed manifold compared to any other. This does not necessarily
imply that a single trajectory touches during its move in time any
equal sized patch in phase space equally often (ergodicity). An
example is the erratic trajectory of a ball on Galton's nail board
which does not touch every nail. Repeated very often the final
distribution is binomial, i.e. ruled by a simple statistical law.
{\em The microcanonical entropy thus measures our ignorance about
the microscopic $6N$-degrees of freedom ($dof$s) needed to
completely determine the system.}

\section{ {\boldmath$S(E)$} is multiply differentiable}

The microcanonical partition sum of N particles interacting via the
two-body potential $U(r-r')$, for simplicity assumed to be bound from below
and confined within the volume ${\cal V}$, is defined as:
\begin{eqnarray}
W(E)&=&\frac{1}{N!(2\pi\hbar)^{3N}}\int_{{\cal V}^N}{d^{3N}\nvec{r}_i
\int{d^{3N}\nvec{p}_i{
\delta\{E-\sum_i^N{\frac{\nvec{p}_i^2}{2m_i}-\Phi[\{\nvec{r}_i\}]}\}}\;\delta^3
\{\sum \nvec{p}_i\}}}\label{micromeg}\\
\Phi[\{\nvec{r}_i\}]&=&\sum_{i<j}{U(\nvec{r}_i-\nvec{r}_j)}=
\frac{1}{2}\sum_{i,j}(U_{i,j} -\delta_{i,j}U_{i,i})\\ W(E)&=&\int_{\cal
V}{d^{3N}\nvec{r}_i~\Theta(E-\Phi[\{\nvec{r}_i\}])\;F_N(E-\Phi[\{\nvec{r}_i\}])}\label{micromegb}\\
F_N(E_{kin}[\{\nvec{r}_i\}])&:=&\frac{1}{N!(2\pi\hbar)^{3N}}\int{d^{3N}\nvec{p}_i\;
\delta(E_{kin}[\{\nvec{r}_i\}]-\sum_i^N{\frac{\nvec{p}_i^2}{2m_i}})\;
\delta^3\{\sum \nvec{p}_i\}}\label{FNkin}\\
&=&\frac{[E_{kin}[\{\nvec{r}_i\}]]^{(3N-5)/2}\prod_1^N{m_i^{3/2}}}
{N!\Gamma(3(N-3)/2)(2\pi\hbar^2)^{3N/2}(2\pi M)^{3/2}}
\end{eqnarray}
The entropy: $S=\ln(W)$ and the inverse temperature
\begin{eqnarray}
\frac{\partial S}{\partial E}=\beta(E)&=&\frac{\partial S}{\partial
E}=\frac{1}{W}\partial W/\partial E\\\nonumber\\& =&\frac{1}{W}\int_{\cal
V}d^{3N}\nvec{r}_i~\Theta(E-\Phi[\{\nvec{r}_i\}])\;F_N(E-\Phi[\{\nvec{r}_i\}])\;
\frac{3N-5}{2(E-\Phi)}\\&=&\left<\Theta(E-\Phi)\frac{3N-5}{2(E-\Phi)}\right>>0
\end{eqnarray}
are {\em non-vanishing and finite} for any finite, non-vanishing
excitation energy ($E>\min{\Phi}$) and $3N>7$.

The kinetic energy per degree of freedom is
\begin{equation}
 \left<\Theta(E-\Phi)\frac{2(E-\Phi)}{3N-5}\right>\ne
T_{thd}=\frac{1}{\beta}=\left<\Theta(E-\Phi)\left[\frac{2(E-\Phi)}{3N-5}\right]^{-1}\right>^{-1}
\end{equation}
The $\Theta$-function in equ.(\ref{micromegb}) excludes the region in the
N-particle configuration space where the total kinetic energy is negative.
By the inclusion of the $\Theta$-function the boundaries in the integral
(\ref{micromegb}) are independent of $E$. For $3N>5$ the
$\delta(E-\Phi[\{r_i\}])$ coming from the differentiation of the
$\Theta$-function is cancelled.

Also
\begin{eqnarray}
\frac{\partial^2S}{\partial
E^2}=\beta'(E)&=&\frac{W''}{W}-\left(\frac{W'}{W}\right)^2\\
&=&\left<\Theta(E-\Phi)\frac{3N-5}{2}\frac{3N-7}{2}\left(\frac{1}{E-\Phi}\right)^2\right>-
\left<\Theta(E-\Phi)\frac{3N-5}{2(E-\Phi)}\right>^2
\end{eqnarray}
is in any case finite.
\section{Consequences} \noindent For really
large $N$, e.g. for a star,
\begin{equation}\frac{d T(E)}{d E}=-\frac{\beta'}{\beta^2}=\frac{1}{cN}\end{equation}
can never be infinite and leading to jumps or loops in $T(E)$ like the ones
claimed by \cite{bouchet03} and \cite{chavanis02b}. Here $c(e)$ is the
microcanonical heat capacity of a star {\em per atom}:
\begin{eqnarray} e&=&E/N\nonumber\\s(e)&=&S(E)/N\nonumber\\
de/dT&=&c(e)\nonumber\\
 c(e)&=&-\frac{(d s/de)^2}{d^2s/de^2}
\end{eqnarray}
As $\beta=\partial S/\partial E> 0$ as well as $\beta'=\partial^2
S/\partial E^2\ne\pm\infty$,
 $c(e)$ can be positive or negative but never $=0$ \cite{gross174}.
  $T(E)$ is a single valued differentiable function at all energies. This
clearly excludes all "hysteretic cyclings", -zones, or "dinosaur's necks"
as introduced by \cite{chavanis02b}. These exotic forms of the
microcanonical caloric curves must be artifacts of the mean-field
approximation for the entropy (equation 5 in ref. \cite{chavanis02b}).

Not only is the microcanonical entropy everywhere multiple differentiable
{\em especially for astronomical numbers of particles}, but also the
interpretation of the entropy as being the {\em geometrical measure} of our
ignorance about the possible initial values of the N-body system is the
most fundamental and by far simplest definition for $S$ of all proposed so
far.
\section{Acknowledgement} I am grateful to Stefano Ruffo for many helpful
discussions during my stay at the Universit\`a di Firenze and INFN, Sezione
di Firenze, via Sansone 1, 50019 Sesto F.no (Firenze), Italy.

\end{document}